\documentclass{hep99}
\usepackage{epsf}
\begin{document}

\title{
QCD factorization for exclusive, non-leptonic $B$ decays
}
\author{\vspace*{-0.2cm}
M Beneke $^\dag$
}

\address{Theory Division, CERN, CH-1211 Geneva 23, Switzerland\\[3pt]
E-mail: {\tt martin.beneke@cern.ch}}

\abstract{Exclusive, non-leptonic, two-body decays of $B$ mesons 
simplify greatly in the heavy quark limit. In this talk I discuss 
the factorized structure that holds in this limit and some of its 
consequences: (a) naive factorization is recovered in a certain 
limit; (b) `non-factorizable' effects are hard and can be 
calculated; (c) strong interaction phases vanish in the heavy quark limit 
(and can be calculated as well). As an illustration, I compute the 
penguin contribution to the decay $B_d\to\pi^+\pi^-$ and its effect 
on the determination of $\sin 2\alpha$.\\
\vspace*{-9.5cm}
\begin{flushright}
CERN-TH/99-319\\
October 1999\\
hep-ph/9910505
\end{flushright}
\vspace*{7.5cm}
} 

\maketitle

\fntext{\dag}{Talk presented at the 
``International Europhysics Conference on High Energy Physics'' (EPS99), 
15-21 July 1999, Tampere, Finland.}

\section{Introduction}

Measuring asymmetries in exclusive decays of $B$ vs. $\bar{B}$ mesons 
is the primary task of four experimental facilities soon in operation.  
While a non-zero asymmetry establishes CP violation unambiguously, the 
interpretation of such a measurement in terms of fundamental theory 
parameters is mostly rather difficult. Hence many strategies that use 
different measurements in an attempt to eliminate or minimize theoretical 
input on strong interaction dynamics, essentially trading theoretical 
uncertainties for experimental ones. It would be clearly useful to 
have some theoretical guidance as well.

In this talk I describe the approach put forward in \cite{BBNS99}. 
The result is that most non-leptonic, two-body $B$ decays can be treated 
as hard processes in QCD and become simple in the heavy quark limit. 
The naive factorization approximation follows as the leading term 
in a systematic expansion. A particularly interesting consequence is 
that soft final state interactions vanish in the heavy quark limit. 
Strong phases are therefore calculable, if the   
form factors and light-cone distribution amplitudes of mesons 
are known.
A simple factorized (in a sense to be explained below) expression 
holds only for the leading term in an expansion in 
$\Lambda_{\rm QCD}/m_b$. Whether the heavy quark limit is an adequate 
approximation for $B$ mesons is quite a different matter, which I do 
not address in this talk. For this reason, all numbers that may 
follow should be 
considered as illustrative of what can be done, once we are convinced 
that the heavy quark limit is good.

\section{The factorization formula}

The basic objects are the matrix elements $\langle M_1 M_2| {\cal O}_i| 
B\rangle$, where $M_1$ and $M_2$ denote the two final state mesons and 
${\cal O}_i$ an operator in the weak effective Hamiltonian. Such matrix 
elements contain both short-distance (`hard') effects, related to 
the large scale $m_b$, and long-distance (`soft', `collinear') effects. 
The idea is to factorize (and compute) hard contributions and to 
parameterize soft and collinear ones, hoping that this leads to some 
simplification. I take the heavy quark limit, i.e. relative corrections 
of order $\Lambda_{\rm QCD}/m_b$ are neglected. Hard and infrared 
contributions will be separated by looking at Feynman diagrams, a method 
familiar from other hard QCD processes (hadron-hadron collisions, 
fragmentation, jets, hard diffraction ...). It is then assumed that 
factorization holds non-perturbatively. This assumption that soft effects 
that may not be visible from Feynman diagrams do not destroy the power 
counting is always implicit in factorization `theorems' and I do not 
know of any example, where it would go wrong -- but it is an assumption.

Before applying these considerations to the case at hand, note that  
two light final state mesons carry energy and momentum $M_B/2$. If all the 
mesons's constituents have large momentum, the meson is appropriately 
described by light-cone distribution amplitudes. There is a finite 
probability for asymmetric partonic fluctuations in which 
a subset of partons carries almost all momentum. For example, for 
a pion with momentum $E$ there is a probability of order $1/E^2$ for 
a $q\bar{q}$ component, in which one of the quarks carries momentum 
of order $\Lambda_{\rm QCD}$. This probability can be estimated from the 
endpoint behaviour of the asymptotic distribution amplitude, since 
the distribution amplitude approaches the asymptotic one in the 
high energy limit. Such endpoint contributions are not appropriately 
described by overlap integrals of light-cone distribution amplitudes. 
If unsuppressed, they require introducing more general non-perturbative 
parameters.

The main results are summarized as follows. Let $M_1$ be the meson that 
absorbs the light spectator quark from the $B$ meson. $M_1$ can be a 
light meson ($M_{M_1}/m_b\to 0$ in the heavy quark limit) or a heavy 
meson ($M_{M_1}/m_b \to\,$finite). The second meson $M_2$ is required 
to be light. Then:

(1) `Non-factorizable' contributions, i.e. contributions that do not 
belong to the $B\to M_1$ form factor or the decay constant $f_{M_2}$ 
of $M_2$, are dominated by hard gluon exchange. This effect 
can be calculated. Since hard gluons transfer large momentum to $M_2$, 
one obtains a convolution with the light-cone distribution amplitude of 
$M_2$ rather than $f_{M_2}$ as in naive factorization. `Non-factorizable' 
soft exchange is suppressed, because the $q\bar{q}$ pair that forms 
$M_2$ is produced as a small colour dipole \cite{BJ}. The cancellation 
of soft gluons is not enough to arrive at a factorization formula. 
It is important that `non-factorizable', collinear gluons cancel as well. 

(2) The $B\to M_1$ form factor is dominated by soft 
gluon exchange for both heavy and light $M_1$ by power counting. 
The reason for this is 
that the $B$ meson contains a soft spectator quark. If $M_1$ is light, 
the endpoint suppression of the light-cone distribution amplitude 
of $M_1$ is 
not sufficient to render the soft contribution power suppressed. In fact, 
the hard gluon correction is suppressed by one power of $\alpha_s$ relative 
to the soft one. If $M_2$ is heavy, the light-cone distribution amplitude 
favours the absorption of a soft quark and the hard contribution is 
suppressed even further. This discussion refers to counting powers. It 
ignores the possibility that resummation of Sudakov logarithms suppresses 
the soft contribution beyond naive power counting. This possibility 
deserves further investigation, even though it appears unrealistic for
realistic $B$ mesons. 

(3) `Non-factorizable', hard gluon exchange between $M_2$ and the 
$B$ meson spectator quark is a leading power effect. Because the gluon 
is hard, the interaction is local in transverse distance and can be 
described by the convolution of three light-cone distribution 
amplitudes. If the `non-factorizable', hard gluon exchange occurs 
between $M_2$ and the other quarks of $B$ and $M_1$, the spectator 
quark can be a distance $1/\Lambda_{\rm QCD}$ away. This implies that 
in this case we must keep the $B\to M_1$ form factor.

(4) Annihilation topologies and higher Fock states of $M_2$ give 
contributions suppressed by powers of $\Lambda_{\rm QCD}/m_b$. 

\begin{figure}[b]
   \vspace{-2.3cm}
   \hspace*{-0.2cm}
   \epsfysize=13.5cm
   \epsfxsize=9cm
   \centerline{\epsffile{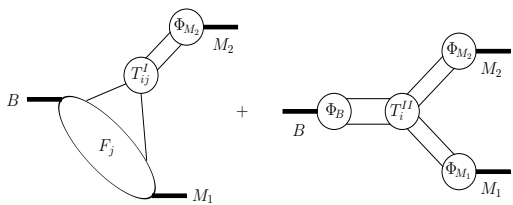}}
   \vspace*{-9.4cm}
\caption[dummy]{\label{fig1}
Graphical representation of (\ref{fff}).}
\end{figure}

The observations collected above are expressed by the factorization 
formula
\begin{eqnarray}
\label{fff}
&&\hspace*{-0.6cm}
\langle M_1(p') M_2(q)|{\cal O}_i|B(p)\rangle =
\nonumber\\
&&\hspace*{-0.2cm}
F_{B\to M_1}(0)\,\int\limits_0^1 \!dx\,T_{i}^I(x)\,\Phi_{M_2}\!(x) 
\\[-0.3cm]
&&\hspace*{-0.2cm}
+\,\int\limits_0^1 \!d\xi dx dy \,T_i^{II}(\xi,x,y)\,
\Phi_B(\xi)\Phi_{M_1}\!(y)\Phi_{M_2}\!(x),
\nonumber
\end{eqnarray} 
where the last line accounts for the hard interaction with the spectator 
quark in the $B$ meson and the equality sign is valid up to corrections 
of order $\Lambda_{\rm QCD}/m_b$. 
$T_i$ denote hard scattering functions that depend 
on longitudinal momentum fractions and $\Phi$ label the light-cone 
distribution amplitudes. See Fig.~\ref{fig1} for a graphical illustration 
of (\ref{fff}). If $M_1$ is heavy, the last line is suppressed 
and should be dropped. In this case an equation like (\ref{fff}) has 
already been used in \cite{PW}. If $M_2$ is a heavy, onium-like meson, 
factorization still holds. If, on the other hand, $M_2$ is a heavy-light 
meson, factorization does not occur. 

The validity of (\ref{fff}) has been demonstrated by an explicit 1-loop 
calculation \cite{BBNS99}. General arguments support factorization to 
all orders in perturbation theory. For a heavy-light final state 
(such as $D\pi$), an explicit demonstration of the cancellation of 
soft and collinear divergences at two loops exists \cite{BBNS}.

One of the consequences of (\ref{fff}) is that in those cases where it 
applies, it also justifies naive factorization as the limit in which 
corrections of order $\alpha_s$ and $\Lambda_{\rm QCD}/m_b$ are 
neglected. The underlying physical picture is that the meson $M_1$ 
is produced in an asymmetric configuration, in which one quark carries 
almost all of the meson's momentum, while the other meson $M_2$ is 
produced in a symmetric $q\bar{q}$ configuration which leaves the 
decay region with little probability for interactions.

The possibility to compute systematically logarithmic corrections to 
naive factorization solves the problem of scheme-dependence often discussed 
in the literature in the context of naive factorization. The hard 
scattering kernels require ultraviolet subtractions related to the 
renormalization of the operator ${\cal O}_i$. If performed consistently, 
these subtractions compensate the scheme- and scale-dependence of the 
Wilson coefficients in the weak effective Hamiltonian.

Another consequence of (\ref{fff}) is that strong 
final state interactions are computable. The imaginary part of the 
amplitude is generated by the imaginary parts of the hard scattering 
amplitudes only. As all other soft effects, soft final state 
interactions are also suppressed by a power of $\Lambda_{\rm QCD}/m_b$ 
and so soft strong phases disappear in the heavy quark limit. 
Since this statement may be a point of controversy \cite{DGPS96}, it 
requires further discussion.

\section{Final state interactions}

When discussing final state interactions, we may choose to use a 
partonic or hadronic language. The partonic language is justified 
by the dominance of hard rescattering in the heavy quark limit. In 
this limit the number of physical intermediate states is arbitrarily 
large. We may then argue on the grounds of parton-hadron duality 
that their average is described well enough (say, up to $\Lambda_{\rm QCD}/
m_b$ corrections) by a partonic calculation. This is the picture 
implied by (\ref{fff}). The hadronic language is in principle 
exact. However, the large number of intermediate states makes 
it almost impossible to observe systematic cancellations, which 
usually occur in an inclusive sum of intermediate states.

To be specific, consider the decay of a $B$ meson into two pions. 
Unitarity implies that 
\begin{equation}
\label{unitarity}
\mbox{Im}\,{\cal A}_{B\to \pi\pi} \sim \sum_n 
{\cal A}_{B\to n}{\cal A}_{n\to \pi\pi}^*.
\end{equation}
The elastic rescattering contribution is related to the $\pi\pi$ 
scattering amplitude, which exhibits Regge behaviour in the high-energy 
($m_b\to\infty$) limit. Hence the soft, elastic rescattering 
phase increases slowly in the heavy quark limit \cite{DGPS96}. On 
general grounds, it is rather improbable that elastic rescattering 
gives an appropriate description in the heavy quark limit. This 
expectation is also borne out in the framework of Regge behaviour, see  
\cite{DGPS96}, where the importance of inelastic rescattering is 
emphasized. However, the approach pursued in \cite{DGPS96} leaves 
open the possibility of soft rescattering phases that do not vanish in 
the heavy quark limit, as well as the possibility of systematic 
cancellations, for which the Regge language does not provide an 
appropriate theoretical framework.

Eq.~(\ref{fff}) implies that such systematic cancellations 
do occur in the sum over all intermediate states $n$. It is worth 
recalling that such cancellations are not uncommon for hard 
processes. Consider the example of $e^+ e^-\to\,$hadrons at large 
energy $q$. While the production of any hadronic final state 
occurs on a time scale of order $1/\Lambda_{\rm QCD}$ (and would 
lead to infrared divergences if we attempted to describe it in 
perturbation theory), the inclusive cross section given by the sum 
over all hadronic final states is described very well by a 
$q\bar{q}$ pair that lives over a short time scale of order $1/q$. In 
close analogy, while each particular hadronic intermediate state 
$n$ in (\ref{unitarity}) cannot be described partonically, the 
sum over all intermediate states is accurately represented by 
a $q\bar{q}$ fluctuation of small transverse size of order $1/m_b$, 
which therefore interacts little with its environment. Note that 
precisely because the $q\bar{q}$ pair is small, the physical picture of 
rescattering is very different from elastic $\pi\pi$ scattering --  
hence the Regge picture is difficult to justify in the heavy quark 
limit. Technically, 2-gluon exchange (plus ladder graphs) between 
a compact $q\bar{q}$ pair with energy of order $m_b$ and transverse 
size of order $1/m_b$ and the other pion does not lead to large 
logarithms and hence no possibility to construct the pomeron. (Notice 
the difference with elastic vector meson production through a virtual 
photon, which also involves a compact $q\bar{q}$ pair. However, 
in this case one considers $s\gg Q^2$ and this implies that the 
$q\bar{q}$ fluctuation is born long before it hits the proton. Hence the 
possibility of pomeron exchange.)

It follows from (\ref{fff}) that the leading strong interaction phase is 
of order $\alpha_s$ in the heavy quark limit. The same statement holds 
for rescattering in general. For instance, 
according to the duality argument, a 
penguin contraction with a charm loop represents 
the sum over all intermediate states of the form 
$D\bar{D}$, $J/\Psi\rho$, 
etc. that rescatter into two pions. 

As is clear from the discussion, parton-hadron duality is crucial for 
the validity of (\ref{fff}) beyond perturbative factorization. Proving 
quantitatively to what accuracy we can expect duality to hold is a yet 
unsolved problem in QCD. Short of a solution, it is worth noting that the 
same (often implicit) assumption is fundamental to 
many successful QCD predictions in 
jet physics, hadron-hadron physics and heavy quark decays.

\section{CP asymmetry in $B_d\to \pi^+\pi^-$ decay}

I now use the factorization formula to compute the time-dependent,  
mixing-induced asymmetry in $B_d\to \pi^+\pi^-$ decay. For reasons 
discussed below I do not consider this result final. However, it 
illustrates the predictivity of a systematic approach to calculate 
non-leptonic decay amplitudes.

\begin{figure}[t]
   \vspace{-2.05cm}
   \hspace*{1cm}
   \epsfysize=20cm
   \epsfxsize=14cm
   \centerline{\epsffile{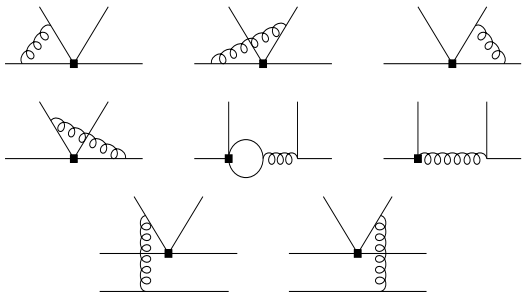}}
   \vspace*{-14.4cm}
\caption[dummy]{\small Order $\alpha_s$ corrections to the hard 
scattering kernels $T^I_i$ (first two rows) and $T^{II}_i$ 
(last row). In the case of $T^I_i$, the spectator quark does 
not participate in the hard interaction and is not drawn. 
The two lines directed upwards represent the two quarks that make up 
$M_2$.
\label{fig2}}
\end{figure}
\begin{figure}[b]
   \vspace{-2.9cm}
   \hspace*{-0.3cm}
   \epsfysize=10cm
   \epsfxsize=7cm
   \centerline{\epsffile{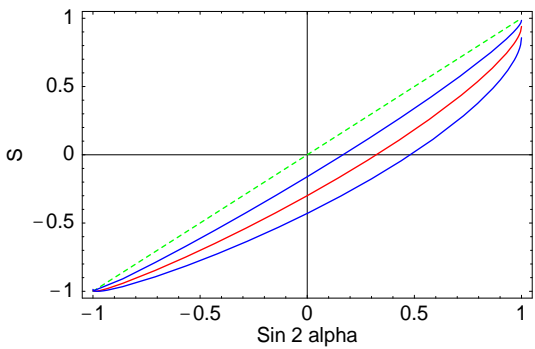}}
   \vspace*{-3cm}
\caption[dummy]{\small Coefficient of $-\sin(\Delta M_{B_d} t)$ vs. 
$\sin 2\alpha$. $\sin 2\beta=0.7$ has been assumed. 
See text for explanation.
\label{alpha}}
\end{figure}

The complete expressions for the decay amplitudes into two pions of 
any charge are given in \cite{BBNS99} including $\alpha_s$ corrections 
to naive factorization. The relevant Feynman diagrams are shown in 
Fig.~\ref{fig2}. In the following I use the same input parameters 
as in \cite{BBNS99}. The time-dependent asymmetry can be expressed as 
\begin{equation}
A(t) = -S\cdot \sin(\Delta M_{B_d}t)+C\cdot\cos(\Delta M_{B_d} t).
\end{equation}
In the absence of a penguin contribution (defined as the contribution 
to the amplitude which does not carry the weak phase $\gamma$ in 
standard phase conventions) $S=\sin 2\alpha$ (where $\alpha$ refers to 
one of the angles of the CKM unitarity triangle) and $C=0$.  
Fig.~\ref{alpha} shows $S$ as a function of $\sin 2\alpha$ with the 
amplitudes computed according to (\ref{fff}). The central of the 
solid lines refers to the heavy quark limit including $\alpha_s$ 
corrections to naive factorization and including the power-suppressed 
term $a_6 r_\chi$ (for notation see \cite{BBNS99}) that is usually 
also kept in naive factorization. The other two solid lines correspond to 
dropping this term or multiplying it by a factor of 2. This exercise 
shows that formally power-suppressed terms can be non-negligible 
(see below), but it also shows that a measurement of $S$ can be 
converted into a range for $\sin 2\alpha$ which may already provide 
a very useful constraint on CP violation.

\section{Outlook}

Much work remains to be done on the theoretical and phenomenological 
side. On the theoretical side, the proof of factorization for a final 
state of two light mesons has to be completed. Then, to define the 
heavy quark limit completely, one must control all logarithms of 
$m_b$. While this is straightforward for the first term on the 
right hand side of (\ref{fff}), this is less trivial for the hard 
scattering term, since $\xi\sim \Lambda_{\rm QCD}/m_b$. This means 
that one must control logarithms of $\xi$ that may appear in higher 
orders. To do this, one has to define the appropriate $B$ meson wave 
function, since $\Phi_B(\xi)$ as it stands still depends on $m_b$. 
Work on these issues is in progress.

Power corrections are an important issue, as $m_b$ is not particularly 
large. There exist `chirally enhanced' corrections that involve the 
formally power suppressed, but numerically large parameter 
$r_\chi=2 m_\pi^2/(m_b (m_u+m_d))$. All terms involving such chiral 
enhancements can be identified, but they involve non-factorizable 
soft gluons. The size of these terms has to be estimated to arrive 
at a realistic phenomenology. 

If this can be done, we expect promising 
constraints and predictions for a large number of non-leptonic, two-body 
final states.

\vspace*{0.3cm}
\noindent
{\bf Acknowledgements.}
It is a pleasure to thank G.~Buchalla, M.~Neubert and C.T.~Sachrajda 
for an on-going collaboration on the subject of this talk. I also wish 
to thank J. Donoghue for interesting discussions. This work was supported 
in part by the EU Fourth Framework Programme
`Training and Mobility of Researchers', Network `Quantum Chromodynamics and
the Deep Structure of Elementary Particles', contract FMRX-CT98-0194 (DG 12
- MIHT).


\begin{thebibliography}{9}
\small
\bibitem{BBNS99} M.~Beneke, G.~Buchalla, M.~Neubert and C.T.~Sachrajda, 
Phys. Rev. Lett. {\bf 83} (1999) 1914.
\bibitem{BJ}
J.D.~Bjorken, Nucl. Phys. (Proc. Suppl.) {\bf B11}, (1989) 325;
M.J.~Dugan and B.~Grinstein, Phys. Lett. {\bf B255} (1991) 583.
\bibitem{PW}
H.D. Politzer and M.B. Wise, Phys. Lett. {\bf B257} (1991) 399.
\bibitem{BBNS} M.~Beneke, G.~Buchalla, M.~Neubert and C.T.~Sachrajda, 
in preparation.
\bibitem{DGPS96}
J.F.~Donoghue, E.~Golowich, A.A.~Petrov and J.M.~Soares,
Phys. Rev. Lett. {\bf 77} (1996) 2178.

\end{thebibliography}
\end{document}